\documentclass[%
 aip,
 amsmath,amssymb,
 reprint,%
]{revtex4-1}

\usepackage{graphicx}% Include figure files
\usepackage{dcolumn}% Align table columns on decimal point
\usepackage{bm}% bold math
\usepackage[utf8]{inputenc}
\usepackage[T1]{fontenc}
\usepackage{mathptmx}
\usepackage{etoolbox}
\usepackage{color}
\newcommand{\tr}[1]{\textcolor{black}{#1}}

\makeatletter
\def\@email#1#2{%
 \endgroup
 \patchcmd{\titleblock@produce}
  {\frontmatter@RRAPformat}
  {\frontmatter@RRAPformat{\produce@RRAP{*#1\href{mailto:#2}{#2}}}\frontmatter@RRAPformat}
  {}{}
}%
\makeatother

\begin{document}

\preprint{AIP/123-QED}

\title[]{An atomistically informed multiscale approach to the intrusion and extrusion of water in hydrophobic nanopores}
% Force line breaks with \\
\author{Gon\c{c}alo Paulo}
\author{Alberto Gubbiotti}%
\author{Alberto Giacomello$^\ast$}
\email{alberto.giacomello@uniroma1.it}
\affiliation{Dipartimento di Ingegneria Meccanica e Aerospaziale, Sapienza Universit\`a di Roma, Rome, Italy}

\date{\today}% It is always \today, today,

\begin{abstract}
Understanding intrusion and extrusion in nanoporous materials is a challenging multiscale problem of utmost importance for applications ranging from energy storage and dissipation to water desalination and hydrophobic gating in ion channels. Including atomistic details in simulations is required to predict the overall behavior of such systems, because the statics and dynamics of these processes depend sensitively on microscopic features of the pore such as the surface hydrophobicity, geometry, and charge distribution and on the composition of the liquid. On the other hand, the transitions between the filled (intruded) and empty (extruded) states are rare events which often require long simulation times difficult to achieve with standard atomistic simulations. In this work, we explored the intrusion and extrusion processes by a multiscale approach in which the atomistic details of the system, extracted from molecular dynamics simulations, inform a simple Langevin model of water intrusion/extrusion in the pore. We then used the Langevin simulations to compute the transition times at different pressures, validating our coarse-grained model by comparing it with nonequilibrium molecular dynamics simulations. %in which pressure was applied to the water surrounding the nanopore, forcing the intrusion/extrusion process. Pressure cycles are performed using the same Langevin model, where the pressure of the system is increased at a given rate until intrusion occurs and reduced at the same rate until extrusion happens. 
The proposed approach reproduces experimentally relevant features such as the time and temperature dependence of the intrusion/extrusion cycles, as well as specific details about the shape of the cycle. This approach also drastically increases the timescales that can be simulated allowing to reduce the gap between simulations and experiments and showing promise for more complex systems.
\end{abstract}

\maketitle

\section{Introduction}

Nucleation of a vapor bubble from a metastable liquid, such as water, is dramatically favored by the presence of inhomogeneities such as dissolved chemical species or the presence of confining surfaces~\cite{crum1979tensile,lefevre2004intrusion,camisasca2020gas}.
The nucleation process in a nanopore can be explained in terms of metastability of the confined vapor bubble (empty state) and of the confined liquid (filled state)~\cite{lafuma2003superhydrophobic,amabili2017collapse,le2022intrusion}.
Transitions between the two states are rare events governed both by external agents such as localized heating~\cite{paul2020single} or pressure~\cite{amabili2017collapse} and by thermal fluctuations.~\cite{gallo2020nucleation}
Continuum deterministic models based on capillary theory are able to predict the stability of the filled/empty states in systems with relatively large (>$10$\,nm) geometrical features~\cite{lefevre2004intrusion,emami2012predicting,checco2014collapse,Amayuelas2023}.
At smaller scales, effects due to extreme confinement, the atomistic nature of matter and thermal fluctuations become increasingly relevant and are thus unavoidable for a quantitative understanding~\cite{guillemot2012activated,tinti2017intrusion,camisasca2020gas,giacomello2020bubble}.

Hydrophobic nanoporous materials immersed in a non wetting liquid can dramatically vary their macroscopic properties depending on what happens at the scale of a single pore -- or even at subnanometer defects within it. 
Predicting under which conditions and how fast liquids fill nanometer-sized cavities and pores or how fast a vapor bubble nucleates is thus a challenging and highly significant multiscale problem, relevant for a number of technological applications. In the context of nanofluidics, the filling state of a nanopore or nanochannel heavily influences its transport properties~\cite{rauscher2008wetting} and hence the capability of controlling the filling and emptying transitions would provide a very promising tool for the design of novel nanofluidic devices \cite{duan2012}.
Nanoscale intrusion is also relevant in other contexts, ranging from the design of systems exploiting porous materials for energy storage and dissipation~\cite{eroshenko2001,fraux2017forced}, to the understanding of hydrophobic gating in biological ion channels~\cite{roth2008bubbles,aryal2015hydrophobic}. 

Here we present a hierarchical multiscale approach which can be used to study the stability and kinetics of atomistic systems in which filling/emptying transitions happen.
We use this protocol to study the dynamics of water intrusion/extrusion a simple cylindrical nanopore characterized by a slightly hydrophobic contact angle ($104^\circ$)~\cite{camisasca2020gas}. 
This system is a simple proxy for a broad category of systems, nanopores,~\cite{lynch2020water} which are crucial for the mentioned applications. Due to the (sub)nanoscale nature of the problem, in which the distribution and orientation of single water molecules is crucial~\cite{tinti2021structure,Paulo2023}, all-atoms simulations are required to fully explore the kinetics of such systems. 
On the other hand, the intrusion and extrusion processes are rare events characterized by long waiting times before a thermally-activated transition is observed; this scenario typically requires long simulation times which are difficult to achieve directly with conventional atomistic simulations, hence calling for specialized rare event techniques~\cite{maragliano2006temperature,bonella2012theory,giacomello2021}. In order to solve this challenge, we adopt a multiscale approach in which a coarse-graining procedure is applied to the all-atom system by identifying a suitable order parameter for the process under investigation, which in this case is the number of water molecules inside the nanopore.
We explored the filling/emptying transition by performing Restrained Molecular Dynamics (RMD)~\cite{maragliano2006temperature}. 
As an output of the simulations, we obtained the free energy as a function of the water filling of the nanopore and the diffusivity~\cite{zhu2012theory} associated with the thermal fluctuations of the same variable at each filling level. 
These quantities were used to build a coarse-grained model for the intrusion/extrusion dynamics based on the Langevin equation. 
Langevin dynamics is a useful technique to study a broad range of nanoscale systems in which thermal fluctuations are relevant~\cite{forrey2007langevin,gubbiotti2019confinement,paquet2015molecular}.
We used Langevin dynamics instead of other possible approaches such as solving Fokker-Planck equation~\cite{van1992stochastic} due to its flexibility, which allowed us to further expand the model to include the effect of time-varying thermodynamic parameters, as pressure, which is crucial to simulate intrusion and extrusion under conditions which are relevant for energy storage and dissipation applications~\cite{fraux2017forced}.

Finally, we validated our coarse-grained model by comparing with nonequilibrium molecular dynamics simulations in which pressure cycles are applied to the nanopore, forcing the filling/drying transition.
Even though the model we obtained refers to a simple cylindrical hydrophobic nanopore, it reproduces features experimentally observed in biological pores, such as stochastic opening and closing~\cite{roth2008bubbles}, and the time-dependence of pressure cycle experiments performed with hydrophobic nanoporous materials~\cite{Grosu2018}.
The same protocol can be used on more complex systems such as zeolites or ion channels, possibly enabling a quantitative prediction for dynamics at experimentally-relevant timescales difficult to achieve directly with all-atom MD.

\section{Methodology}
The rare event nature of the filling/emptying transitions calls for a reduced description of the dynamics. This involves choosing a suitable collective variable and coarse-graining procedure. 
Here we use the pore filling $N$, \textit{i.e.} the number of water molecules inside the nanopore, as the order parameter for the intrusion/extrusion process. 
\tr{We assume that the process can be described by the Fokker-Planck equation
\begin{equation}\label{eq:FokkerPlanck}
    \frac{\partial P(N,t)}{\partial t} = 
        \frac{\partial}{\partial N}\left(
            \beta D(N)\frac{\mathrm{d}F}{\mathrm{d}N}P(N,t)
            +D(N)\frac{\partial P(N,t)}{\partial N}
        \right)\;,
\end{equation}
where $P(N,t)$ is the time-dependent probability distribution,}
$D(N)$ is the diffusivity of the pore filling, 
$\beta=1/k_B T$ is the inverse thermal energy and $F(N)$ is the free energy of the system, \textit{i.e.}, a function such that the equilibrium distribution is given by
\begin{equation}\label{eq:equilibrium}
P(N)\propto e^{-\beta F(N)}\,.
\end{equation}
\tr{The It\^{o} process associated to Eq.~\eqref{eq:FokkerPlanck} is described by the Langevin equation}
\begin{equation}
    \label{eq:langevin}
    \frac{\mathrm{d}N}{\mathrm{d}t}=-\beta D\frac{\mathrm{d}F}{\mathrm{d}N}+\frac{\mathrm{d}D}{\mathrm{d}N}+g(N)\xi(t)\,,
\end{equation}
where $g(N)$ determines the intensity of the thermal noise and $\xi(t)$ is a white noise process.
\tr{
    The second term on the right hand side of Eq.~\eqref{eq:langevin} is needed to cancel out the extra term generated in the Fokker-Planck equation by the fact that the diffusivity $D(N)$ is state-dependent ~\cite{lau2007state,gubbiotti2019confinement}
}.
The mobility $M(N)=\beta D(N)$ and the noise intensity $g(N)$ are linked via the fluctuation-dissipation relation
\begin{equation}
M(N)=\beta D(N)=\frac{\beta g(N)^2}{2}\,.
\end{equation}

It is important to point out that both $F$ and $D$ depend on $N$, and that a dependence on other external parameters such as pressure $P$ and temperature $T$ can be considered in more refined models for those quantities.
The state-dependent diffusivity is shown to be crucial for the correct modeling of the process at hand, which could be further improved by considering the inertia of the coarse-grained variable, as opposed to the overdamped model here presented. 

\subsection{Molecular dynamics setup}

\begin{figure}
    \centering
    \includegraphics[width=0.5\textwidth]{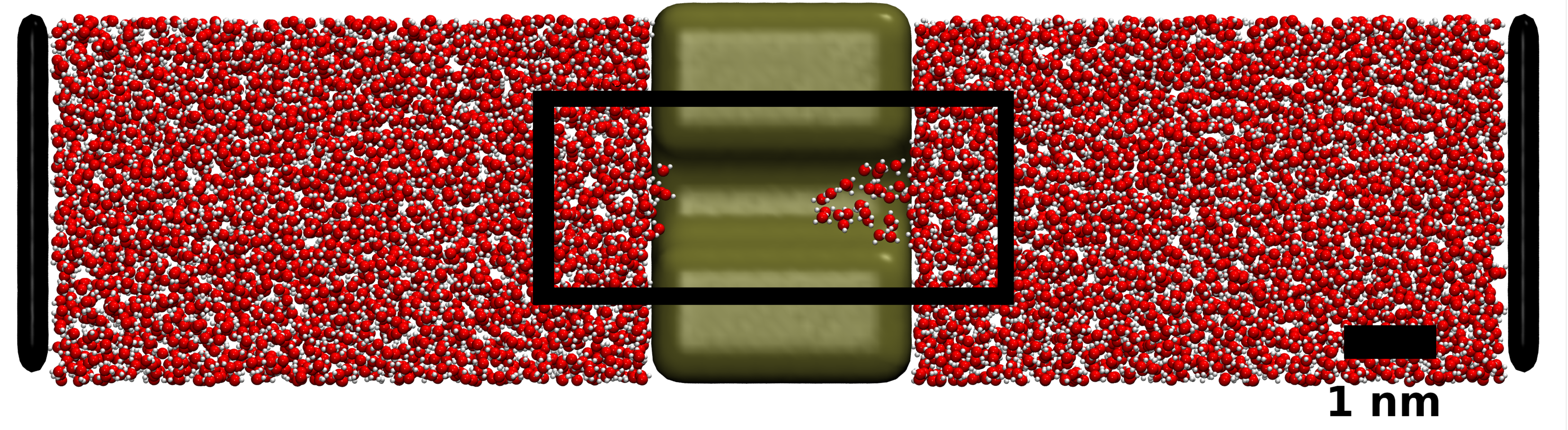}
	\caption{System used in this study, consisting of two reservoirs of liquid SPC/E water separated by a cylindrical nanopore formed by Lennard-Jones atoms. 
 %The diameter of the nanopore is $1.4\,$nm and its length is $2.8\,$nm, while the contact angle is $104^\circ$. 
 The pistons, represented in black, are used to control the pressure by applying a suitable force. The black rectangle enclosing the pore represents the box used to compute the water filling $N$. This figure was realized with the VMD software~\cite{vmd}.}
    \label{fig:system}
\end{figure}

The system studied via atomistic simulations is described in the following and represented in Fig.~\ref{fig:system}. 
The aim of the MD simulations is to extract the two quantities $F$ and $D$ which define the coarse-grained dynamics in Eq.~\eqref{eq:langevin}. 
The membrane is made of a slab of fixed atoms in fcc arrangement, with lattice spacing 0.35 $nm$, from which a nanopore was excavated. 
Water (SPC/E~\cite{berendsen1987missing}) was placed on both sides of the slab. 
The interaction of the water with the surface was tuned~\cite{camisasca2020gas} so that the contact angle is $104^\circ$. The nanopore has  diameter of 1 $nm$ and a  length of 2.8 $nm$. 
\tr{We have chosen the specific model of water, dimensions of the pore, and contact angle to simulate narrow hydrophobic nanopores in contact with water at a reasonable computational cost. Other contact angles, pore dimensions, and models of water could be used in conjunction with the presented techniques. Pores of this particular contact angle and similar sizes have been previously simulated \cite{Paulo2023} as a proxy to study real hydrophobic nanoporous materials. These dimensions are also in the range of biological nanopores affected by hydrophobic gating allowing to tackle wetting on both classes of problems.}
At each end of the water reservoirs a thin slab of hydrophilic material is present, which is used as a piston to control the pressure of the system \cite{marchio2018}. 
The NVT ensemble was sampled using a Nos\'e--Hoover chains thermostat \cite{martyna1992} at 310$^\circ$ K with a chain length of 3.

\subsection{Free energy computation}

We use Restrained Molecular Dynamics (RMD)~\cite{maragliano2006temperature} to compute the free energy as a function of the pore filling.
This is done by adding a harmonic restraint to the original Hamiltonian of the system,
\begin{equation}\label{eq:restraint}
    H_N(\boldsymbol{r},\boldsymbol{p})=H_0(\boldsymbol{r},\boldsymbol{p})+\frac{k}{2}\left(N-\tilde{N}(\boldsymbol{r})\right)^2\,,
\end{equation}
where $\boldsymbol{r}$ and $\boldsymbol{p}$ are the positions and momenta of all the atoms, respectively, $H_0$ is the unrestrained Hamiltonian, $k$ is a harmonic constant which was set to $1$\,kcal/mol, $N$ is the desired number of water molecules in a box centered around the nanopore, and $\tilde{N}$ is the related counter,
\begin{equation}    \tilde{N}(\boldsymbol{r})=\sum_{i=0}^{N_{ox}}\prod_{\alpha=1}^3\Theta_\alpha(r_{i\alpha})\;,
\end{equation} 
where the index $i$ spans all the oxygen atoms of the water molecules and the index $\alpha$ spans the three directions in space with a smoothed version of the indicator function reading 
\begin{equation} \Theta_\alpha(r_{i\alpha}) = \frac{1}{\exp\left[-\lambda(h_\alpha+r_{i\alpha})\right]+1}-\frac{1}{\exp\left[\lambda(h_\alpha-r_{i\alpha})\right]+1}\;,
\end{equation} 
where the parameter $\lambda$ tunes the degree of smoothing and $h_\alpha$ are the (half) box sizes in each dimension. 
In our simulations, $\lambda=3$ \AA$^{-1}$, $h_1=h_2=11$ \AA,  $h_3=23.125$ \AA, where direction 3 is aligned to the pore axis while directions 1 and 2 are perpendicular to it.

Independent MD simulations  were performed for several values of $N$, which allowed us to computed the gradient of the free energy from the mean force \cite{maragliano2006string}
\begin{equation}
    \frac{\partial F}{\partial N} =  k\left\langle (N-\tilde{N}(\boldsymbol{r})) \right\rangle_N\;,
\end{equation}
where $\left\langle...\right\rangle_N$ represents the expected value in a simulation run with the Hamiltonian $H_N$. The free energy profile was then reconstructed by simple integration.

\subsection{Diffusivity computation}

\begin{figure*}
    \centering
    \includegraphics[width=\linewidth]{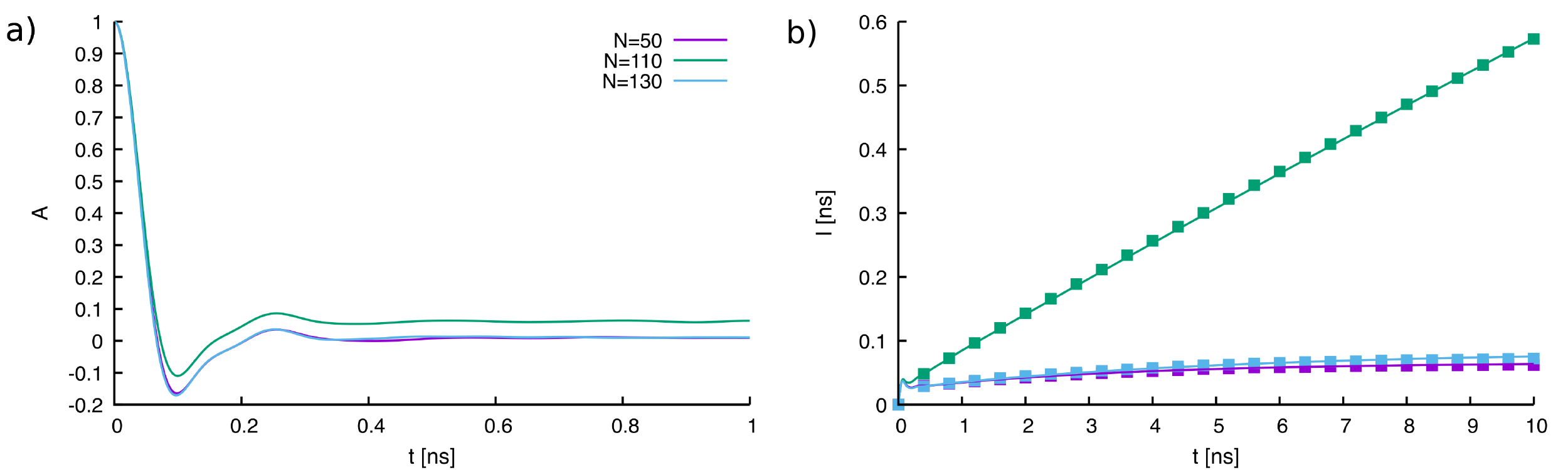}
    \caption{a) Decay of autocorrelation function of three the pore filling., in three relevant cases: $N=50$ (empty minimum), $N=110$ (maximum) and $N=130$ (filled minimum). b) Integral of the autocorrelation for the same three cases. The points are the numerically computed values, while the colored lines represent the exponential fitting, showing excellent agreement after a time of $~0.5$ ns. }
    \label{fig:diffusivity}
\end{figure*}

The diffusivity associated to the filling dynamics can be computed with the same restrained simulations described in the previous section, computing the autocorrelation integral of $N$. This approach has been previously used for computing position-dependent particle diffusivity~\cite{zhu2012theory,nagai2020position}. Here we use it to compute the state-dependent diffusivity $D(N)$ associated with a coarse-grained variable such as the filling of the nanopore. 

\tr{
    The main idea is that we can use the same Langevin Equation, Eq.~\ref{eq:langevin}, also for the restrained system, under the assumption of strong restraining forces. This allows us to consider the harmonic restraint as the only relevant source of drift, and also to neglect the dependency of diffusivity on filling. The advantage of this approach is clear if we consider that a single set of simulations can be used to compute both the free energy and the diffusivity, and that unbiased approaches, such as the evaluation of the slope of the mean squared displacement would not allow us to consider the state dependence of the diffusivity.
}

\tr{
    The Langevin equation corresponding to a RMD simulation with the restraint described in Eq.~\eqref{eq:restraint} is, for a sufficiently large value of the constant $k$,
    \begin{equation}
        \Delta\dot{N}=-k\beta D(N)\Delta N+\sqrt{2D(N)}\xi\;,
    \end{equation}
    where $N$ is the value at which the filling is restrained, and $\Delta N$ is a shifted filling variable such that $\left\langle\Delta N\right\rangle=0$.
    The variance of $\Delta N$ is
    \begin{equation}\label{eq:variance}
        \left\langle\left(\Delta N\right)^2\right\rangle=\frac{1}{k\beta}
    \end{equation}
    and the autocorrelation
    \begin{equation}
        \left\langle \Delta N(t)\Delta N(0)\right\rangle
            = \left\langle \Delta N(0)\Delta N(0)\right\rangle e^{-k\beta Dt}\;.
    \end{equation}
    By integration, we get 
    \begin{equation}\label{eq:autocorr_int}
        \int\limits_{0}^{\infty}\left\langle \Delta N(t)\Delta N(0)\right\rangle
            = \frac{1}{k^2\beta^2D}\;.
    \end{equation}
    We can then compute the local diffusivity from RMD simulations using Eq.~\eqref{eq:variance} and Eq.~\eqref{eq:autocorr_int}, which provide a value independent of $k$, as
}
\begin{equation}
        D(N)=\frac{\left\langle \Delta N^2 \right\rangle_N^2}{\int_0^\infty \left\langle \Delta N(t)\Delta N(0)\right\rangle_N\mathrm{d}t}\,.
        \label{eq:diffusivity}
\end{equation}

\tr{
    Defining the normalized autocorrelation function 
    \begin{equation}
        A_N(t) = \frac{\left\langle \Delta \tilde N(t)\Delta \tilde N(0)\right\rangle_N}{\left\langle \Delta N^2 \right\rangle_N}\;,
    \end{equation}
    we have
    \begin{equation}
        D(N) = \frac{\left\langle \Delta N^2 \right\rangle_N}{\int_0^\infty A_N(t) \mathrm{d}t}\;.
    \end{equation}
    If the autocorrelation integral can be assumed to be weakly dependent on temperature, the main temperature dependence of the diffusivity can be ascribed to the variance of $\Delta N$, hence resulting in a diffusivity proportional to the temperature.
}
Three examples of autocorrelation function $A_N(t)$ are shown in Fig.~\ref{fig:diffusivity}a, for the pore fillings corresponding to the two minima and the maximum of the free energy. 
It can be observed that, at the top of the free energy barrier, the decay of autocorrelation is much slower, resulting in a much smaller diffusivity. Even if the initial part of the autocorrelation decay (the first $~0.5$ ns) shows a complex behavior which can be ascribed to the inertia of the water molecules, after that time the decay is essentially exponential as is shown in fig.~\ref{fig:diffusivity}b, where the autocorrelation integrals $I(t)=\int_0^t A(t')\mathrm{d}t'$ are shown together with the corresponding exponential fittings.
The limit value of the exponential fitting for $t\to\infty$ has been taken as the value of the autocorrelation integral at the denominator of Eq.~\eqref{eq:diffusivity}.

\subsection{Langevin dynamics simulations}

The dynamics of $N$ according to Eq.~\eqref{eq:langevin} is specified giving the free energy $F$ and the diffusivity $D$, which can be extracted from MD simulations.
We used an overdamped Langevin code~\cite{gubbiotti2019confinement} to integrate Eq.~\eqref{eq:langevin} using the Euler-Maruyama algorithm~\cite{miguel2000stochastic}. \tr{These simulations are employed to be able to compute some observables, like the times of intrusion and extrusion of the nanopore, which are too expensive to compute with standard molecular dynamics simulations.}
Another approach to compute the intrusion/extrusion times would have been to exploit rate theory \cite{Kramers1940,Hnggi1990} fed with the free energy profile $F(N,P)$ and the state dependent diffusivity $D(N)$:
\begin{equation}
    t_{rt} = \int_{N_a}^{N_b}\frac{e^{\beta F(N,P)}}{D(N,P)}dN \int_{N_a}^{N_b}e^{-\beta F(N,P)}\phi(N,P) dN,
    \label{eq:ratetheory}
\end{equation}
with $t_{rt}$ being the mean drying/wetting times and $N_a$ and $N_b$  the filling levels of the initial and final states, respectively and where $phi(N)$, the committor, is given by:
\begin{equation}
    \phi(N,P) = \int_{N_a}^{N}\frac{e^{\beta F(N',P)}}{D(N',P)}dN'\left({\int_{N_a}^{N_b}\frac{e^{\beta F(N',P)}}{D(N',P)}}dN'\right)^{-1},
    \label{eq:comittor}
\end{equation}
Using  $N_a$ as the dry state and $N_b$ as the filled state, yields the wetting time, while using $N_a$ as the wet state and $N_b$ as the dry state yields the drying time.  \tr{The dependence of the free energy profile and diffusivity on the pressure is discussed in the next sections}.

\section{Results}
\subsection{Free energy and diffusivity}

\begin{figure*}
    \centering
    \includegraphics[width=\linewidth]{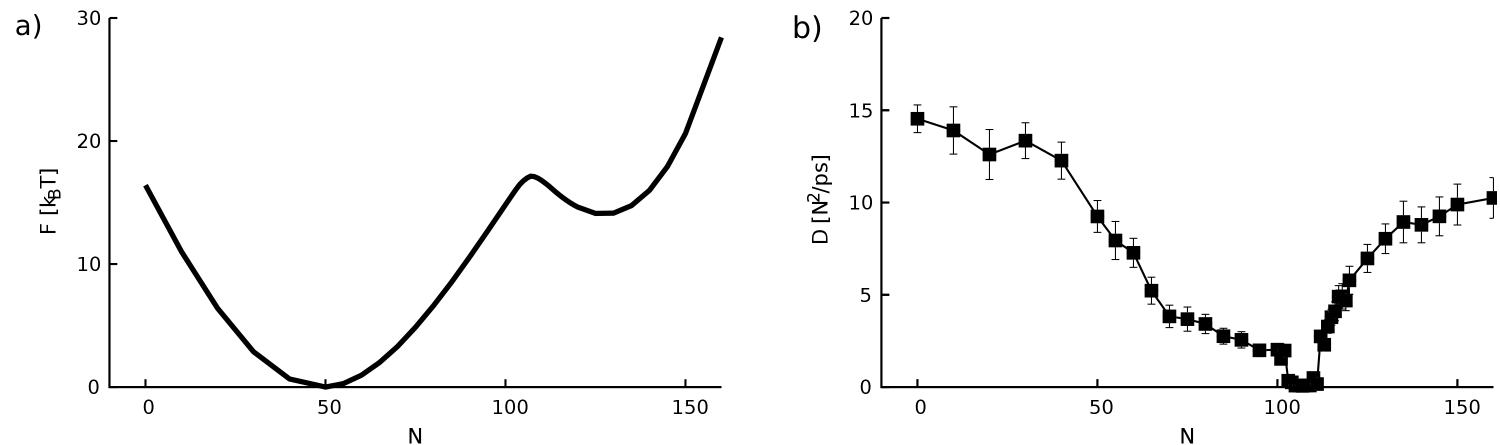}
    \caption{a) Free energy as a function of the water filling. The empty and filled states are separated by barriers of 18 and 3 $k_BT$ respectively. At ambient pressure the pore is empty more than 99\% of the time. The error bars are comparable with the width of the line. b) Diffusivity as a function of the water filling, decreasing for smaller levels of filling, until it reaches a minimum at the filling value associated with the maximum of the barrier, and increasing again when the pore is filled.}
    \label{fig:free_energy}
\end{figure*}

The free energy $F$ as a function of the pore filling $N$ is shown in Fig.~\ref{fig:free_energy}a as computed from RMD simulations. $F$ has a global minimum at low filling levels, corresponding to the empty state, and a local minimum at higher filling levels, corresponding to the filled state. For $N < 110$ the pore contains a vapor bubble whose size grows as $N$ decreases. \tr{We can estimate the size of the critical bubble associated with extrusion by considering the difference between the filling level in the filled state, 130 water molecules, and at the maximum of the barrier, 110 water molecules, which corresponds to a volume of ca. 0.6 $nm^3$, 1/4 of the total pore volume.}

The intrusion barrier of this system, i.e., the barrier that needs to be overcome for the system go from the empty state to the filled state, $F_w\approx 18\;k_
BT$ and the extrusion barrier $ F_d\approx 3$~$k_BT$. 
This means that while it is possible that the system spontaneously fill, the time that will take for it to empty is $\exp(F_w- F_d)\approx 10^6$ times faster meaning that the filled state is barely observable at ambient conditions. This also means that, for this system, it is very hard to estimate precisely the intrusion times using standard MD techniques, because one would have to collect statistics of an extremely improbable transition.

We compute the diffusivity $D(N)$ via Eq.~\eqref{eq:diffusivity} by evaluating the correlation time of the water fluctuations inside the pore at different filling levels along RMD simulations; more details of these calculations are found in the supplementary material.
Figure~\ref{fig:free_energy}b shows the result of this computation.  We observe that the diffusivity associated to nucleation events is non-monotonically dependent on $N$, decreasing as $N$ increases from smaller filling levels, or equivalently, increasing with the size of the vapor bubble, as also observed for bulk bubble cavitation \cite{menzl2016molecular}.
After reaching a minimum corresponding to the maximum of the free energy, the diffusivity increases again for higher fillings.

\subsection{Effect of applied pressure}

\begin{figure*}
    \centering
    \includegraphics[width=\linewidth]{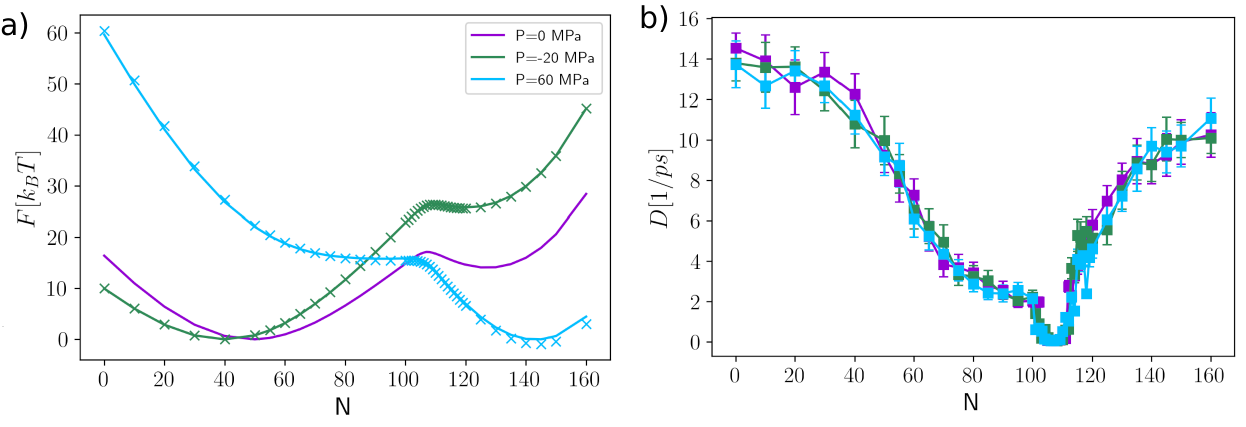}
    \caption{Free energy and diffusivity at different pressures. The tilting method (symbols) for the free energy gives indistinguishable results to the simulations with an explicitly pressure applied (solid lines). In panel a) the solid line represents the measured profiled and the symbols represent the tilting of Eq.~\eqref{eq:tilt} starting from the profile at 0 MPa. \tr{Panel b) shows that} there is no significant change in the trend of the diffusivity or its absolute value for the three different pressures tested.}
    \label{fig:different_pressures}
\end{figure*}

It was previously shown ~\cite{tinti2017intrusion,giacomello2020bubble} that applying a pressure to the system will add to the free energy a term proportional to the volume of liquid in the pore, $V$ 
\begin{equation}
F(N,P)=F(N,0)-PV(N) \;,
\label{eq:tilt}
\end{equation}
corresponding to a tilting of the free energy profile with respect to $N$. The dependence $V(N)$ is simply given by $V=N \rho$, with $\rho$ the density of water inside the pore.
The assumption~\eqref{eq:tilt} can be exploited  to obtain the free energy in a broad range of pressures without repeating the simulations. We validated this hypothesis by explicitly performing RMD simulations at $P=60$~MPa and $-10$~MPa, see 
Fig.~\ref{fig:different_pressures}a.

The diffusivity of the collective variable was also computed at different pressures and no major difference was observed within the tested pressure range, see Fig.~\ref{fig:different_pressures}b. These assumptions allowed us to compute the rates via Langevin simulations for a broad range of pressures using only the outcome of the RMD simulations at $P=0$~MPa.

\subsection{Validation of the Langevin Model}
\subsubsection{Computation of the intrusion and extrusion times}

\begin{figure}
    \centering
    \includegraphics[width=1\linewidth]{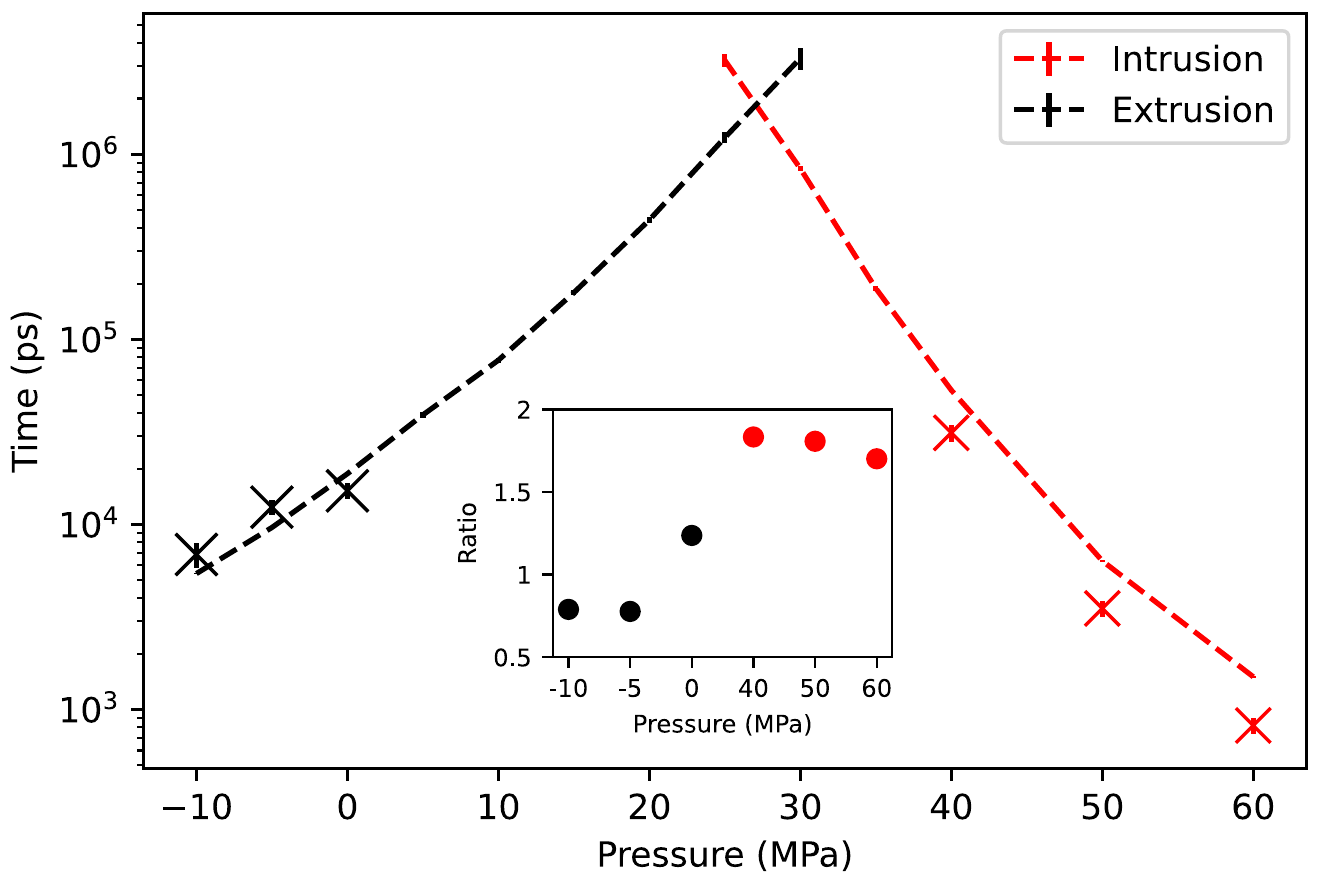}
    \caption{The estimated extrusion (black) and intrusion (red) times at different pressures. Due to the computational cost, the intrusion times  extracted from molecular dynamics (MD) (crosses) were only computed for $P=40$, $50$, and $60$~MPa and the extrusion (black) times for $P=-10$, $-5$, and $0$~MPa. The predicted times using Langevin Dinamics (LD) are the dashed lines, capturing well the trends and the values computed using MD. In the inset we show the ratio between the LD times and the MD times. For intrusion, the times computed using LD are at the biggest deviation 2x the ones computed using MD, and for extrusion the agreement is better, with the times computed using LD being at biggest deviation 0.8x smaller. }
    \label{fig:rates}
\end{figure}

As mentioned previously the filling/emptying transitions are rare events, and as such, the intrusion and extrusion times are difficult to compute directly by standard molecular dynamics simulations. High pressure simulations can be used to compute the intrusion times, as the barrier associated with intrusion is lowered at higher pressures. Similarly, low pressure simulations can be used to compute the extrusion rates, as low pressures lead to low extrusion barriers. We compute the intrusion times directly using MD for $P=40$, $50$, and $60$ MPa and the extrusion times $P=0$, $-5$, and $-10$ MPa. We use these reference times to validate our coarse-grained Langevin approach, which considerably extends the times that can be computed, see Fig.~\ref{fig:rates}.

\tr{We computed the first passage time distributions $\rho_{W}(t)$ and $\rho_{D}(t)$ of the filling/emptying transitions by initializing the system in the filled/empty state and stopping the simulation when the other state was reached, while recording the times required for the transitions.}

Langevin dynamics simulations show a good agreement with the trends observed using MD. The intrusion times computed using LD are at most 2x the time computed using MD for all pressure values, see inset Fig.~\ref{fig:rates}, and the extrusion times computed with LD are at the least 0.8x the time computed using MD. For comparison, in the nucleation literature order-of-magnitude matching of the rates is considered satisfactory \cite{Wang2008,Valeriani2005}. Part of the error on the determination of the rates may be due to the fact that the distribution of intrusion and extrusion times can be considered a exponential distribution for long times, and the values we sample with MD are skewed for lower times. The distribution of intrusion and extrusion times is discussed further in supplementary note 1.

The agreement between MD and LD simulations suggest that LD can be used to simulate intrusion/extrusion times in pressure ranges inaccessible with standard MD, especially when considering that the rates can be calculated without performing additional MD simulations after the initial calculation of the free energy profile and diffusivity at $P=0$. 
For the considered nanopore, LD allowed us to compute the intrusion times at $P<40$~MPa and extrusion times at $P>0$~MPa, which is up to 3 orders of magnitude higher than the ones that we could simulate with MD. In particular, LD also gives access to the coexistence pressure, where the intrusion and extrusion times are equal. The coexistence pressure for our model pore is ca. 28 MPa  with typical times in the $\mu s$ range, see Fig. \ref{fig:rates}. This means that observations longer than some $\mu s$, at pressures higher than 28 MPa, will find the pore intruded most of the time. 

It is important to note that the coexistence pressure estimated by the rates is lower than the figure predicted by the Young-Laplace equation, which is normally used to compute the intrusion pressure in nanopores \cite{giacomello2020bubble}:
\begin{equation}
    \Delta P = -\frac{2 \gamma_{lv} \cos\theta}{r} \; ,
    \label{eq:laplace}
\end{equation}
with $\gamma_{lv}$ the surface tension of the intruding liquid, $\theta$ its contact angle with the pore material, and $r$ the radius of the pore. Using the nominal radius of our model nanopore, 0.7 nm, this equation yields $\Delta P \approx 44$ MPa, while if one uses the water accessible radius of 0.55 nm, it yields  $\Delta P\approx 55$ MPa. This value is close to the one  at which the empty state is no longer stable, i.e., 60 MPa, see Fig.~\ref{fig:free_energy}a. 
Indeed, the Young-Laplace equation predicts not the coexistence pressure, where the two states have equal probability, but the spinodal pressure, where the empty state is no longer metastable. For long nanopores, where the intrusion barriers are very large, typically > $50 k_BT$, Eq.~\eqref{eq:laplace} correctly predicts the intrusion pressure, as the pores remain trapped in the empty state up to the pressure where the barrier is sufficiently low.  On the other hand, short pores, as the one considered here, have smaller intrusion barriers such that intrusion starts before the pressure predicted by Eq.~(\ref{eq:laplace}).

\subsubsection{Intrusion/extrusion under pressure cycles}
%\tr{Revised up to here AG}

We have assumed so far that the dynamics of the filling variable are overdamped and Markovian. 
Under these assumptions, the proposed multiscale approach (atomistically informed Langevin simulations) can be used to perform simulations where external parameters change in time. In the case of porous materials, it is of direct experimental interest to study filling-emptying transitions under cycles of increasing and decreasing pressure, also known as intrusion-extrusion cycles.

\begin{figure}
    \centering
    \includegraphics[width=1\linewidth]{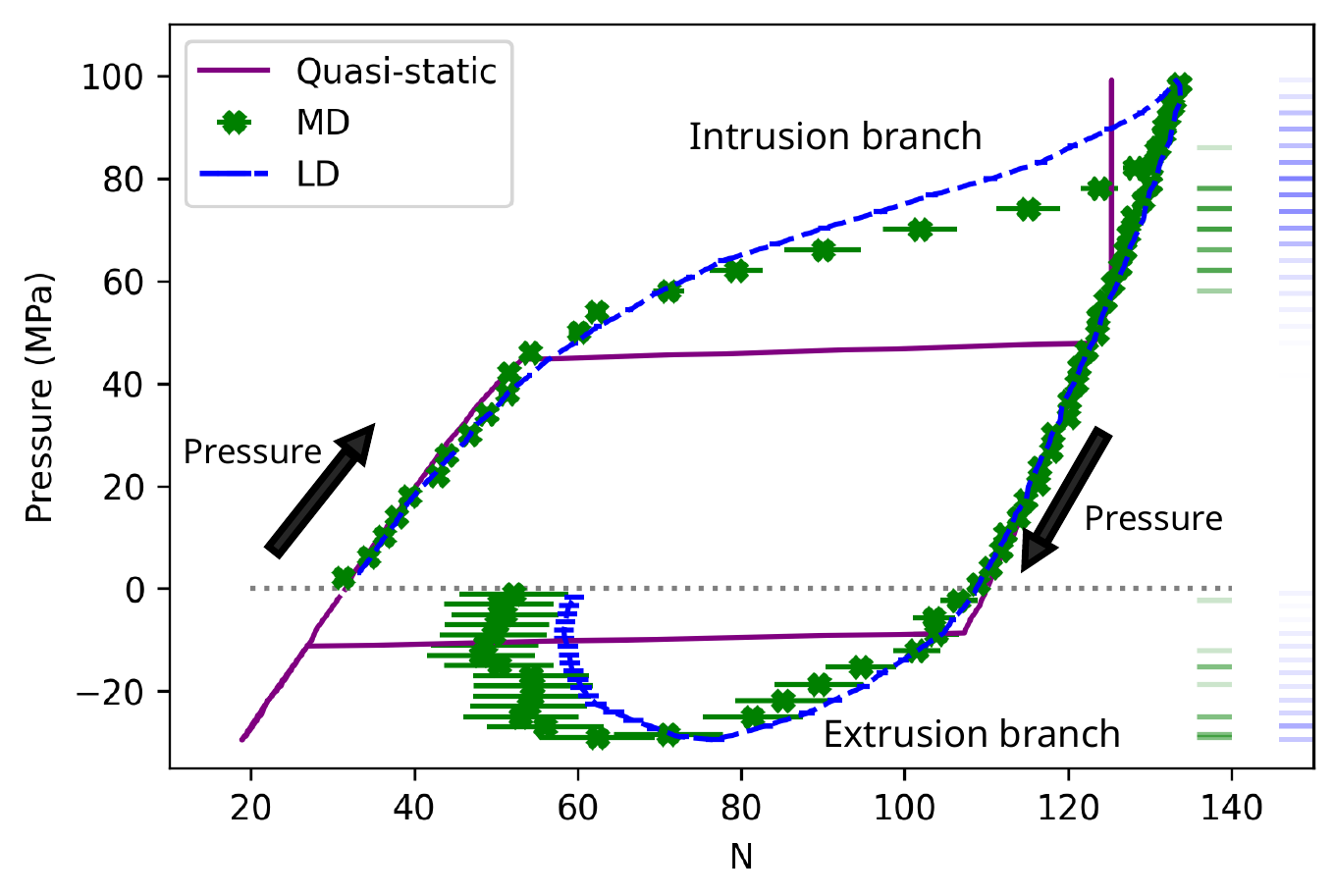}
    \caption{Intrusion and extrusion cycles using molecular dynamics and Langevin dynamics. This figure shows a $8$~ns pressure cycle computed by both LD (blue), MD (green), and considering a quasi-static approximation (purple). The dotted line is a guide to the eyes and represents the 0 pressure line. The MD cycle is an average of 30 cycles while the LD cycle is the average of 1000 cycles. The quasi-static approximation uses eq.~\eqref{eq:ratetheory} for estimating the rates. To capture the slopes during compression/decompression we use the values of the water filling for the filled/empty states at that pressure. The blue (LD) and green lines on the side of the (MD) represent individual intrusion and extrusion events. 
    }
    \label{fig:pressurecycles}
\end{figure}

Figure~\ref{fig:pressurecycles} shows that LD successfully predicts the behavior observed in non-equilibrium molecular dynamics. The cycle starts with the pore dry and no applied pressure. Then pressure is continuously increased, with number of waters in the pore region slowly increasing, corresponding to the pre-intrusion branch. After a certain pressure is reached water intrudes into the pore and N increases very fast, corresponding to the intrusion branch. After the pressure reaches the maximum of 100 MPa it is decreased at the same rate, with the filling of the pore slowly diminishing, corresponding to the pre-extrusion branch. After a critical pressure the pore can extrude, which we call the extrusion branch. We finish the cycle by decreasing the pressure down to -20 MPa and then increasing it again up to 0 MPa, with some pores still extruding at this time. 
 LD is able to capture the increase in the number of water molecules in the pore region as pressure is increased before intrusion, as well as the decrease of water molecules inside the pore before extrusion; both these behaviors are related with the compression and decompression of water before intrusion and extrusion actually take place. LD also captures the fact that intrusion and extrusion are stochastic events %and that, given the typical barriers of the pore we consider, 
that can occur at different pressures for different realizations, discussed in more detail in supplementary note 2. Stochasticity emerges from the relatively small transition times characterizing our system (Fig.~\ref{fig:rates}), which, on average give rise to a slope in the intrusion and extrusion branches, as opposed to  a sharp transition that would characterize a deterministic event as that described by Eq.~\eqref{eq:laplace}. 

\tr{Because the cycle in Fig.~\ref{fig:pressurecycles}  is performed in a finite and short time, not all the pores are extruded when the pressure is decreased, leading to a cycle with partial extrusion.  Both MD and LD capture this behavior. We note that longer cycle times, like the ones shown in Fig.~\ref{fig:comparison}, do not show partial extrusion.}

Langevin dynamics simulations provide an estimate of the intrusion pressure that departs from MD by ca. 5-10 MPa, consistently with the estimated times of intrusion which are higher than the ones measured using MD (Fig.~\ref{fig:comparison}). This may be due to the dynamics of intrusion and extrusion not being truly overdamped and Markovian, like we assume, or due to inaccuracies in the determination of the free energy profile $F(N)$ or of the state-dependent diffusivity $D(N)$. 

In Fig.~\ref{fig:pressurecycles}, a quasi-static approximation of the pressure cycle is also reported, as used in other works\cite{tinti2017intrusion}. 
\tr{This approximation considers that the pressure cycle is constituted of constant pressure steps and that intrusion/extrusion happens if the average intrusion/extrusion time is faster than the duration of the step itself. This is equivalent to hypothesizing that, for a certain cycling time, intrusion and extrusion will happen as soon as their respective barriers are sufficiently small. Differently from the previous use of this approximation \cite{tinti2017intrusion}, which resorted to an Arrhenius like law to compute the intrusion and extrusion times, we used eq.~\eqref{eq:ratetheory}, which more precisely estimates the actual rates (Fig. \ref{fig:rates}).
Figure \ref{fig:pressurecycles}  shows that the quasi-static approximation underestimates the absolute value of both the intrusion and the extrusion pressures  and does not capture the slopes observed during intrusion/extrusion. Both discrepancies can only be described by considering the stochastic nature of the process, which makes the intrusion and extrusion processes progressive, rather than all-or-none. The fundamental flaw of this approximation is thus considering that intrusion and extrusion always happen at a prescribed value of the barrier. We further highlight that the quasi-static approximation would not normally capture the slopes observed in the figure when the empty and filled states are compressed/decompressed, which were added by taking the relevant minimum of the free energy, at any given pressure. }

The comparisons in Fig.~\ref{fig:pressurecycles} demonstrate that Langevin dynamics can capture the transition dynamics  for extremely fast cycles ($8$~ns). Note that the rates at which pressure is changed in these cycles is of the order of 20 MPa/ns, while in  typical experiments cycles are in the range of MPa/s \cite{lefevre2004intrusion,Grosu2018,Johnson2023,Picard2021}. This 9 order of magnitude difference makes it hard to extrapolate from MD cycles to behavior that is observed in experiments. The fact that LD can make cycles on the order of MPa/ms allows it to be used to explain experimentally relevant phenomena such as the dependence of the intrusion and extrusion pressures on the cycle time and temperature, as shown in the next section.

\subsection{Using Langevin dynamics to approach experimental timescales}

\begin{figure*}
    \centering
    \includegraphics[width=1\linewidth]{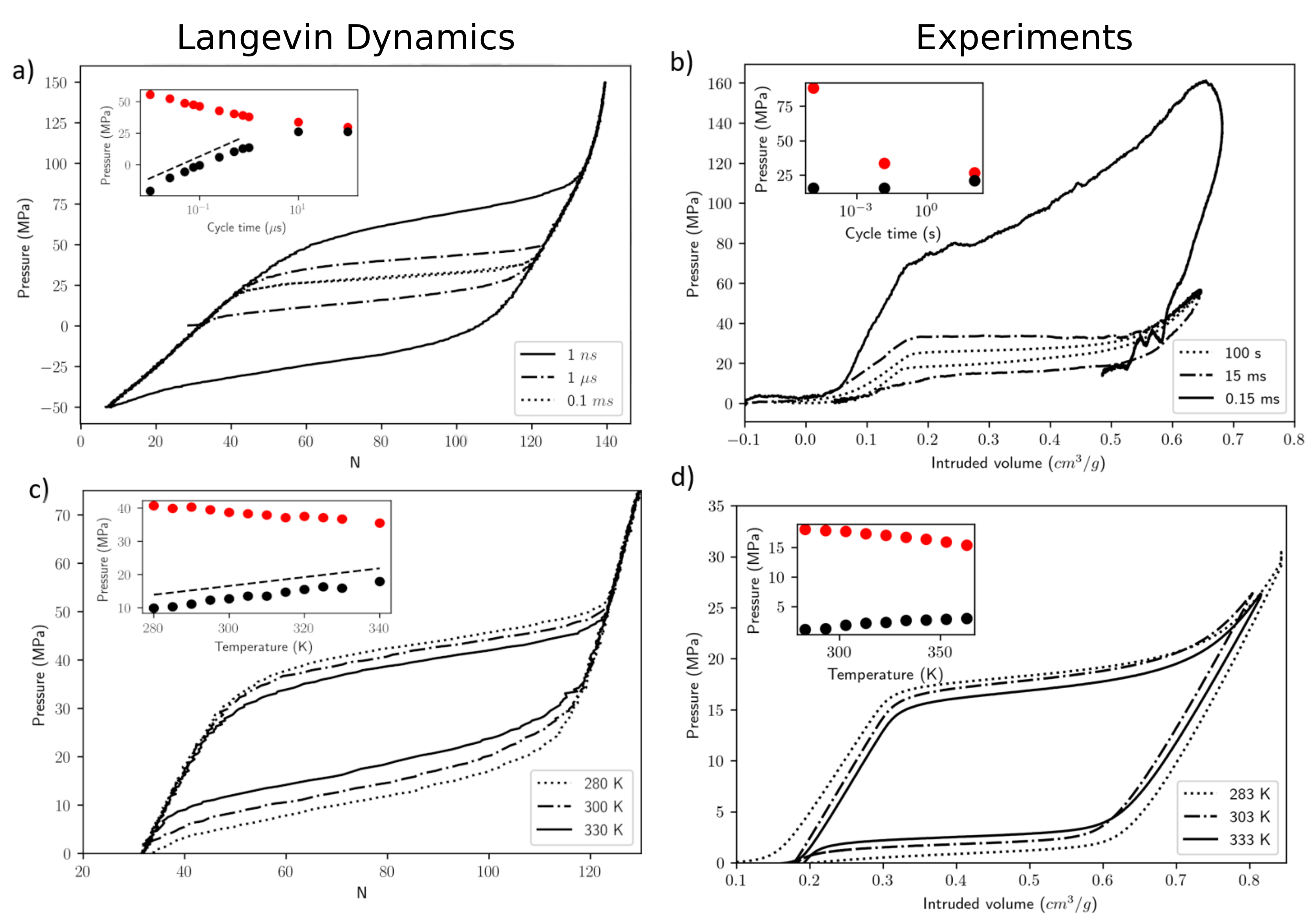}
    \caption{
    Intrusion/extrusion cycles for different cycle times at constant temperature: simulation results at 310K and spanning 4 orders of magnitude of cycle time (a), compared with shock experiments\cite{Sun2021} in ZIF-8 (b). 
    Slower cycles show less pronounced intrusion and extrusion slopes, lower intrusion pressures, and higher extrusion pressures, thus resulting in a higher hysteresis. The insets in both panels show how intrusion and extrusion pressures get closer for slower cycles. \tr{The black dashed like corresponds to a linear fit of the extrusion pressures, whose slope is given by $k_BT/V_c$ \cite{guillemot2012activated}, with $V_c$ being the critical volume for extrusion; in this case $V_c=0.58$.}
    Panel (c) shows the predicted temperature dependence of 10 $\mu$s pressure cycles using LD, compared to experiments on the silica-gel C8 \cite{Grosu2014} (d). Higher temperatures favor both intrusion and extrusion, lowering hysteresis, as shown in the insets, which show data points taken from the original work. \tr{In the inset to panel (c), a linear fit of the extrusion pressure is used estimate the line tension $\tau = -10.7$~pN.}
    }
    \label{fig:comparison}
\end{figure*}

The coarse-grained LD dynamics can be exploited to explore trends in the intrusion/extrusion cycles varying different experimentally relevant parameters, like the temperature or the cycle time. In this section we discuss general effects and compare them with the experimental behavior. We do not try, and cannot, match the observed results with the ones observed in individual materials, as this would require a computation of the diffusivity and free energy of comparable systems. However, we show that the general trends are captured for a variety of materials and conditions, pointing to a common physical explanation.

In Fig.~\ref{fig:comparison}, we compare experimental intrusion/extrusion cycles \cite{Grosu2014,Sun2021} with LD simulations to highlight the behavior that the proposed multiscale approach is able to capture. 
Both the intrusion and the extrusion pressure depend on the cycle time\cite{tinti2017intrusion}, see Fig.~\ref{fig:comparison}a, with longer cycle times continuously reducing the intrusion pressure and increasing the extrusion pressure; accordingly the hysteresis of the cycle continuously decreases until it tends to zero as the thermodynamic limit is approached.
\tr{The slope of the extrusion branch is associated with the volume of the critical bubble required for nucleation \cite{guillemot2012activated}, and the estimated value, 0.58 $nm^3$, is close to the value of 0.60 $nm^3$ we estimate from the free energy profile.}
Fast cycles can have much larger intrusion and extrusion pressures, as reported in recent shock experiments\cite{Sun2021} in Fig.~\ref{fig:comparison}b. As with the case of temperature, the intrusion and extrusion slopes depend on the cycle time, with faster cycles having a higher spread of intrusion/extrusion pressures, and the spread being asymmetric for the two processes. This behavior should be taken into consideration when performing porosity measurements.
The slowest cycles performed using LD is of 0.1 ms, matching the fastest experimental cycle in shock experiments in Fig.~\ref{fig:comparison}b. We remark that we were able to perform simulations of cycle times spanning at least 5 orders of magnitude, getting us closer to make direct predictions of the experimental behavior.

\tr{In Fig.~\ref{fig:comparison}c we compute via LD the dependence of intrusion and extrusion pressures on the temperature, assuming that, over the considered interval, the main effect comes from increasing the thermal energy $k_BT$ available to the system and the diffusivity $D$, according to $D\propto T$.} Previous experiments on silica-gel C8 \cite{Grosu2014} are shown in Fig.~\ref{fig:comparison}d for comparison. This trend confirms that both intrusion and extrusion are thermally-activated events, which are accelerated by increasing the temperature of the system. The change in the  intrusion pressure, however, is smaller than that in the extrusion pressure. This is due to the fact that, at least in cylindrical pores, intrusion barriers are more sensitive to pressure than extrusion ones \cite{tinti2017intrusion}. This effect is further enhanced by the pore length, which explains why for slender pores it is sometimes considered that only extrusion is thermally activated \cite{lefevre2004intrusion,guillemot2012activated}. 

\tr{One could estimate the line tension $\tau$ of the system by fitting the temperature-dependent extrusion pressures via a macroscopic nucleation theory\cite{guillemot2012activated}}
\begin{equation}
    P_\mathrm{ext}(T)= \frac{k_BT}{V_c}\ln \left(\frac{t_0}{t_\mathrm{pre}}\right) +\tau C_1 + C_2 
\end{equation}
\tr{where $t_0$ is the cycle time,  $t_\mathrm{pre}$ the prefactor found in rate equations, and $C_1$ and $C_2$ depend on $\theta$ and on $r$ and are given in the original reference\citep{guillemot2012activated}. Because we do not explicitly include in the calculations of Fig.~\ref{fig:comparison}c the temperature dependence of $\theta$, $\gamma_{lv}$, and $\tau$, only two parameters are needed in the fit: $\tau$ and $t_\mathrm{pre}$. %This is slightly different from what was done previously where the prefactor is estimated and a temperature dependence is obtained. 
In our case, the leading temperature dependence of $P_\mathrm{ext}$ is expected from the term $k_BT$; in principle, also $t_\mathrm{pre}$ could introduce a logarithmic dependence, because it is related to the diffusivity \cite{Hnggi1990}. However, data in Fig.~\ref{fig:comparison}c are fitted satisfactorily by a straight line. 
We obtain $\tau=-10.7$~pN, a value in excellent agreement with previous estimates for similar systems \cite{tinti2017intrusion}. The prefactor $t_\mathrm{pre}$ is estimated to be ca. 4350 ps, which is on the same order of magnitude of the timescale of 700 ps estimated from Fig.~\ref{fig:rates}. %, but orders of magnitude longer with previous microscopic scale  of previous results in the literature \cite{tinti2017intrusion,guillemot2012activated,lefevre2004intrusion}. 
In the future, it would be interesting to explicitly take into account the temperature dependence of the contact angle, surface tension, and line tension. This study, however, would require new simulations at different temperatures, which are beyond the scope of the present work.}

We finally remark that the cycles in Fig.~\ref{fig:comparison} display a slope in both the intrusion and the extrusion branches. In liquid porosimetry\cite{LenyLen1998,Giesche2006} this behavior that is typically interpreted as originating from a distribution of pore radii and the derivative $\mathrm d V/\mathrm d P$ is used for  measuring it assuming that pores intrude only at a specific pressure, $\Delta P$, which is related to the radius by Laplace equation~\eqref{eq:laplace}, \tr{or to the presence of defects}. %Observing a slope in the intrusion pressures is generally interpreted as a distribution of pore radii.
Actually, our results show that slopes in the intrusion and extrusion branches of a $P(V)$ isotherm can emerge as a result of the stochastic nature of the intrusion/extrusion processes, even when the pore radius is monodisperse as in our system. In other words, at any given pressure there is a distribution of transition times, which means that not all pores in the experimental sample will fill/empty at the same pressure, unless the experiment is significantly slower than the typical time required to switch between states (Fig.~\ref{fig:rates}).

\section{Conclusions}
In this work we used a multiscale approach consisting of atomistically informed Langevin simulations to study the intrusion and extrusion of water in a simple hydrophobic nanopore. The approach consists in computing the free energy profile and the diffusivity associated with the transition between the filled and empty states via all-atom molecular dynamics; these quantities are then used in a coarse-grained (Langevin) simulations with only one degree of freedom. %\tr{Alternatively, one could also use formulations evolving the probability density function.}
The proposed framework allowed us to simulate much longer timescales than the ones accessible to molecular dynamics and to study intrusion-extrusion transitions under pressure cycles; both these improvements bring simulations closer to experimental conditions. 

The model was validated in two different ways; first, by computing the intrusion and extrusion times at high and low pressures using molecular dynamics simulations as references, and comparing them to those estimated using Langevin dynamics. Langevin dynamics showed a very close agreement with direct molecular dynamics simulation. Secondly intrusion-extrusion cycles induced by a time-varying pressure were also computed using Langevin dynamics, which was found to accurately predict the behavior observed in non-equilibrium molecular dynamics.

Because Langevin dynamics significantly extends the timescale amenable to simulations by 5 orders of magnitude (from tens of ns to hundreds of $\mu$s), we explored as a proof of principle the role of the cycle time and temperature on the behavior of intrusion/extrusion cycles. Results compare well with the trends of available experiments: intrusion pressure decreases with higher temperatures and longer cycle times and extrusion pressure has the opposite trend. \tr{Fitting the latter results with a nucleation theory provided estimates for the critical bubble volume and for the line tension in excellent agreement with our own free-energy profiles and with previous simulations\cite{tinti2017intrusion}, respectively. } Interestingly, the intrusion and extrusion branches had a slope in the $P(V)$ curves for most cycle times and temperatures, \tr{which are normally interpreted as coming from a distribution of pore radii or from the presence of defects; however, since we considered a single pore radius, our data demonstrate that the stochastic nature of the intrusion and extrusion processes are sufficient to produce such slopes}.

Overall, the presented results suggest that atomistically informed Langevin dynamics can be a useful tool for simulating intrusion/extrusion phenomena that are not within reach of  standard molecular dynamics, decreasing the gap between simulations and experiments. The same procedure can be used for more complex systems, e.g., realistic models of mesopores or zeolites and for biological ion channels.

\section{Acknowledgement}

This research is part of a project that has received funding from the European Research Council (ERC) under the European Union's Horizon 2020 research and innovation programme (grant agreement No. 803213). The authors acknowledge PRACE for awarding us access to Marconi100 at CINECA. % (Project NanoGAS).

\section*{Data Availability Statement}

The data that support the findings of this study are openly available in Zenodo at http://doi.org/10.5281/zenodo.7871060.

\section*{Supplementary Material}
See supplementary material for discussions on the distribution of intrusion and extrusion times at constant pressure and the distribution of intrusion and extrusion pressure in pressure cycles.

\appendix

\nocite{*}
\bibliography{aipsamp}% Produces the bibliography via BibTeX.

\end{document}